\documentclass{article}
\usepackage{ismir,amsmath,cite,url}
\usepackage{graphicx}
\usepackage{color}
\usepackage{xcolor}
\usepackage{float}
\title{Randomly weighted CNN\MakeLowercase{s} for (music) audio classification}

\twoauthors
  {Jordi Pons} {Music Technology Group\\ Universitat Pomepeu Fabra, Barcelona}
  {Xavier Serra} {Music Technology Group\\ Universitat Pomepeu Fabra, Barcelona}

\begin{document}

\maketitle
\vspace{-7mm}
\begin{abstract}
	\vspace{-2mm}
The computer vision literature shows that randomly weighted neural networks perform reasonably as feature extractors. Following this idea,
we study how non-trained (randomly weighted) convolutional neural networks perform as feature extractors for (music) audio classification tasks.  
We use features extracted from the embeddings of deep architectures as input to a classifier~-- with the goal to compare classification accuracies when using different randomly weighted architectures. By following this methodology, we run a comprehensive evaluation of the current deep architectures for audio classification, and provide evidence that the architectures alone are an important piece for resolving (music) audio problems using deep neural networks.
\vspace{-3mm}
\end{abstract}

\section{MOTIVATION -- FROM PREVIOUS WORKS}
\label{sec:motivation}

Some intriguing properties of deep neural networks are periodically showing up in the scientific literature. Examples of those are: \textit{(i)} perceptually non-relevant signal perturbations that dramatically affect the predictions of an image classifier \cite{goodfellow2014explaining,szegedy2013intriguing}; \textit{(ii)} although there is no guarantee of converging to a global minima that might generalize, image classification models perform well with unseen data \cite{krizhevsky2012imagenet,he2015delving}; or \textit{(iii)}~non-trained deep neural networks are able to perform reasonably well as image feature extractors \cite{saxe2011random,ulyanov2017deep,rosenfeld2018intriguing}. 
In this work, we exploit one of the above listed properties \textit{(iii)} to evaluate how discriminative deep audio architectures are before training.

\begin{figure*}[h]
	\centerline{
		\includegraphics[width=0.85\linewidth]{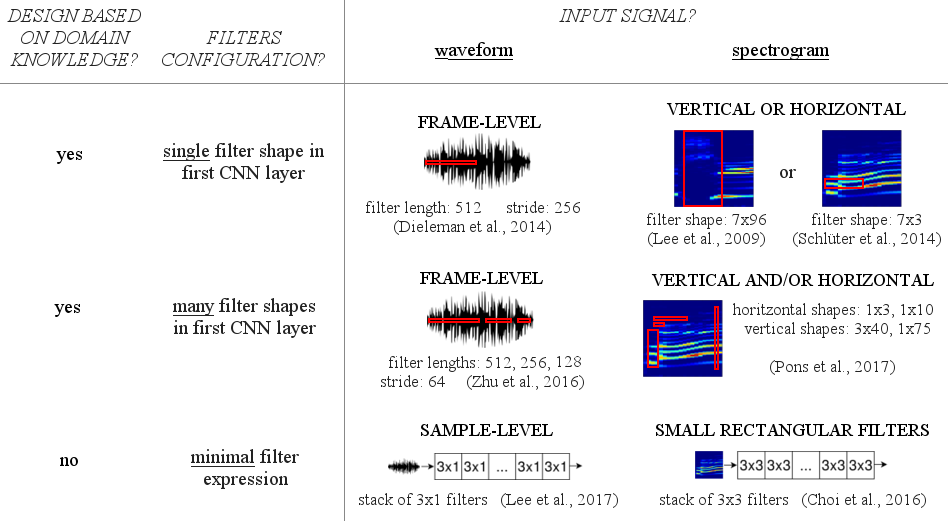}}
	\vspace{-3mm}
	\caption{CNN front-ends for audio classification tasks -- with examples of possible configurations for every paradigm.}
		\vspace{-4mm}
	\label{fig:sota}
\end{figure*}
\noindent Previous works already explored the idea of empirically studying the qualities of non-trained (randomly weighted) networks, but mainly in the computer vision field:

$\mathbf{\cdot}$ Saxe et al. \cite{saxe2011random} studied how discriminative are the architectures themselves by evaluating the classification performance of SVMs fed with features extracted from non-trained (random) CNNs.\footnote{CNNs stands for Convolutional Neural Networks.}They showed that a surprising fraction of the performance in deep image classifiers can be attributed to the architecture alone. Therefore, the key to good performance lies not only on improving the learning algorithms but also in searching for the
most suitable architectures. Further, they showed that the (classification) performance delivered by random CNN
features is correlated with the results of their end-to-end trained
counterparts -- this result means, in practice, that one can bypass the time-consuming process of learning for evaluating a given architecture. We build on top of this result to evaluate current CNN architectures for audio classification.

$\mathbf{\cdot}$ Rosenfeld and Tsotsos \cite{rosenfeld2018intriguing} fixed most of the model's weights to be random, and only allowed a small portion of them to be learned. By following this methodology, they showed a small decrease in image classification performance when these models were compared to their fully trained counterparts. Further, the performance of their non fully-trained models can be summarized as follows: 

\noindent\hspace{0mm}\textit{DenseNet}\cite{huang2017densely} $\gg$ \textit{ResNet}\cite{he2016deep} $>$ \textit{VGG}\cite{simonyan2014very} $\gg$ \textit{AlexNet}\cite{krizhevsky2012imagenet}

\noindent What matches previous works reporting how these (fully trained) models perform\cite{huang2017densely,he2016deep,simonyan2014very}, confirming the performance correlation between randomly weighted models and their trained counterparts found by Saxe~et~al.~\cite{saxe2011random}

$\mathbf{\cdot}$ Adebayo et al. \cite{adebayo2018local} empirically assessed the local explanations of deep image classifiers to find that randomly weighted models produce explanations similar to those produced by models with learned weights. They conclude that the architectures introduce a strong prior which affects the learned (and not learned) representations.

$\mathbf{\cdot}$ Ulyanov et al. \cite{ulyanov2017deep} also showed that the structure of a network (the non-trained architecture) is sufficient to capture useful features for the tasks of image denoising, super-resolution and inpainting. They think of any designed architecture as a \textit{hand-crafted} model where prior information is embedded in the structure of the network itself. This way of thinking resonates with the rationale behind the family of audio models designed considering domain knowledge (see section~\ref{sec:arch}) -- what denotes that in both audio and image fields it exists the interest of bringing together the end-to-end learning literature and previous research.

Few related works exist in the audio field -- and every randomly weighted neural network we found in the audio literature was a mere baseline~\cite{choi2017transfer,kim2018one,arandjelovic2017look}.
Inspired by previous computer vision works, we study which audio architectures work the best via evaluating how non-trained CNNs perform as feature extractors.
To this end, we use the CNNs' embeddings to construct feature vectors for a classifier -- with the goal to compare classification performances when different randomly weighted architectures are used to extract features.
To the best of our knowledge, this is the first comprehensive evaluation of randomly weighted CNNs for (music) audio classification.
\newpage
Extreme learning machines (ELMs) \cite{schmidt1992feedforward,pao1994learning,huang2006extreme} and echo state networks (ESNs) \cite{jaeger2001echo} are also closely related to our work. In short, ELMs are classification/regression models\footnote{Support Vector Machines are also classification/regression models.}that are based on a single-layer feed-forward neural network with random weights. They work as follows: first, ELMs randomly project the input into a latent space; and then, learn how to predict the output via a least-square fit. More formally, we aim to predict:
\vspace{-1mm}
\[\hat{Y} = W_2 \hspace{1mm} \sigma( W_1 X ),\]

\noindent where $W_1$ is the (randomly weighted) matrix of input-to-hidden-layer weights, $\sigma$ is the non-linearity, $W_2$ is the matrix of hidden-to-output-layer weights, and \textit{X} represents the input. The training algorithm is as follows: \textit{1)}~set $W_1$ with random values; \textit{2)} estimate $W_2$ via a least-squares fit:
\vspace{-3mm}
\[W_2 = \sigma( W_1 X )^+ Y\]
\noindent where $^+$ denotes the Moore-Penrose inverse. Since no iterative process is required for learning the weights, training is faster than stochastic gradient descent~\cite{huang2006extreme}.
Provided that we process audio signals with randomly weighted CNNs, ELM-based classifiers are a natural choice for our study -- so that all the pipeline (except the last layer) is based on random projections that are only constrained by the structure of the neural network. Although ELMs are not widely used by the audio community, they have been used for speech emotion recognition\cite{han2014speech, kaya2016combining}, or for music audio classification\cite{scardapane2013music, loh2006elm, khoo2012automatic}. ESNs differ from ELMs in that their random projections use recurrent connections. Given that the audio models we aim to study are not recurrent, we leave for future work using ESNs -- see \cite{scardapane2017semi, blackboxgeorg} for audio applications of ESNs.

\vspace{-3mm}
\section{ARCHITECTURES}
\label{sec:arch}
\begin{figure}[h]
	\centering
	\includegraphics[width=1\linewidth]{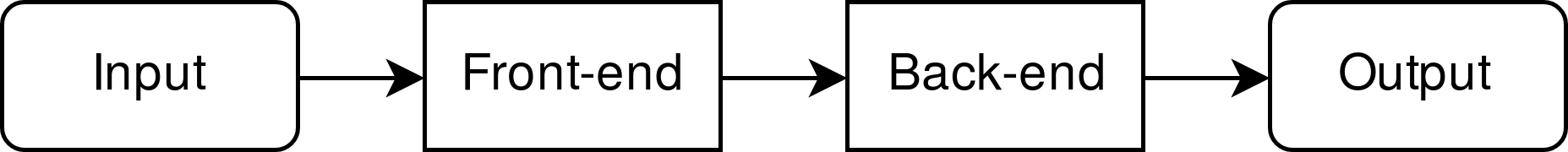}
	\vspace{-6mm}
	\caption{The deep learning pipeline.}
	\label{fig:parts_top}
\end{figure}

In this work we evaluate the most used deep learning architectures for (music) audio classification. In order to facilitate the discussion around these architectures, we divide the deep learning pipeline into two parts: front-end and back-end, see Figure~\ref{fig:parts_top}. The front-end is the part that interacts with the input signal in order to map it into a latent-space, and the back-end predicts the output given the representation obtained by the front-end. 
Note that one can interpret the front-end as a ``feature extractor" and the back-end as a ``classifier". Given that we compare how several non-trained (random) CNNs perform as feature extractors, and we will use out-of-the-box classifiers to predict the classes: this literature review focuses in introducing the main deep learning front-ends for audio classification.

\vspace{-2mm}

\textbf{\textit{Front-ends}} --- 
These are generally conformed by CNNs \cite{dieleman2014end,choi2016automatic,zhu2016learning,pons2017designing,pons2017timbre}, since these can encode efficient representations by sharing weights\footnote{Which constitute the (eventually learnt) feature representations.}along the signal. 
Figure \ref{fig:sota} depicts six different CNN front-end paradigms, which can be divided into two groups depending on the used input signal: waveforms \cite{dieleman2014end,zhu2016learning,lee2017sample} or spectrograms \cite{choi2016automatic,pons2017designing,pons2017timbre}.
Further, the design of the filters can be either based on domain knowledge or not. For example, one leverages domain knowledge when
the \textit{frame-level single-shape}\footnote{Italicized names correspond to the front-end types in Figure \ref{fig:sota}.} front-end for waveforms is designed so that 
the length of the filter is set to be the same as the window length in a STFT \cite{dieleman2014end}. Or for a spectrogram front-end, it is used \textit{vertical} filters to learn spectral representations~\cite{lee2009unsupervised} \textit{or~horizontal} filters to learn longer temporal cues~\cite{schluter2014improved}. 
Generally, a single filter shape is used in the first CNN layer \cite{dieleman2014end,choi2016automatic,lee2009unsupervised,schluter2014improved}, but some recent work reported performance gains when using several filter shapes in the first layer \cite{zhu2016learning,pons2017designing,pons2017timbre,phan2016robust,pons2016experimenting,chenhigh}. Using many filters promotes a more rich feature extraction in the first layer, and facilitates leveraging domain knowledge for designing the filters' shape. 
For example: a \textit{frame-level many-shapes} front-end for waveforms can be motivated from a multi-resolution time-frequency transform\footnote{The Constant-Q Transform \cite{brown1991calculation} is an example of such transform.}perspective -- with window sizes varying inversely with frequency \cite{zhu2016learning}; or since it is known that some patterns in spectrograms are occurring at different time-frequency scales, one can intuitively incorporate \textit{many vertical and/or horizontal} filters to efficiently capture those in a spectrogram front-end \cite{pons2017designing,pons2017timbre,pons2016experimenting,phan2016robust}.
As seen, using domain knowledge when designing the models allows to naturally connect the deep learning literature with previous relevant signal processing work.
On the other hand, when domain knowledge is not used, it is common to employ a deep stack of small filters, e.g.: 3$\times$1 in the \textit{sample-level} front-end used for waveforms~\cite{lee2017sample,van2016wavenet,rethage2017wavenet}, or 3$\times$3 in the \textit{small-rectangular filters} front-end used for spectrograms~\cite{choi2016automatic}. These VGG-like\footnote{VGG: a computer vision model based on a deep stack of 3$\times$3 filters.}models make minimal assumptions over the local stationarities of the signal, so that any structure can be learnt via hierarchically combining small-context representations. 

\vspace{-1mm}

\section{METHOD}

Our goal is to study which CNN front-ends work best via evaluating how non-trained models perform as feature extractors.
Our evaluation is based on the traditional pipeline of \textit{features extraction} + \textit{classifier}.
We use the embeddings of non-trained (random) CNNs as features: for every audio clip, we compute the average of each feature map \linebreak (in~every~layer) and concatenate these values to construct a feature vector\cite{choi2017transfer}.
The baseline feature vector is constructed from 20~MFCCs, their $\Delta$s and $\Delta\Delta$s. We compute their mean and standard deviation through time, and the resulting feature vector is of size 120. We set the widely used MFCCs + SVM setup as baseline. 
To allow a fair comparison with the baseline, CNN models have $\approx$~120 feature maps~-- so that the resulting feature vectors have a similar size as the MFCCs vector. Further, we evaluate an alternative configuration with more feature maps ($\approx$3500) to show the potential of this approach. Model's description omit the number of filters per layer for simplicity -- full implementation details are accessible online.\footnotemark[8]

\vspace{-2mm}
\subsection{Features: randomly weighted CNNs}
\vspace{-1mm}
Except for the \textit{VGG} model that uses ELUs as non-linearities~\cite{choi2016automatic,clevert2015fast}, the rest use ReLUs~\cite{glorot2011deep} -- and we do not use batch normalization, see discussion in section~\ref{sec:repro}. 
We use waveforms and spectrograms as input to our CNNs: 

\vspace{2mm}

\noindent\textbf{\textit{Waveform inputs}} --- are of $\approx$ 29sec (350,000 samples at 12kHz) and the following architectures are under study:

$\cdot$ \textit{\underline{Sample-level:}} is based on a stack of 7 blocks that are composed by a 1D-CNN layer (filter length: 3, stride:~1) and a max-pool layer (size: 3, stride: 3) -- with the exception of the input block which has no max-pooling and its 1D-CNN layer has a stride of 3 \cite{lee2017sample}. 
Averages to construct the feature vector are computed after every pooling layer, except for the first layer that are computed after the CNN. 

\begin{figure}[h]
	\centering
	\includegraphics[width=1\linewidth]{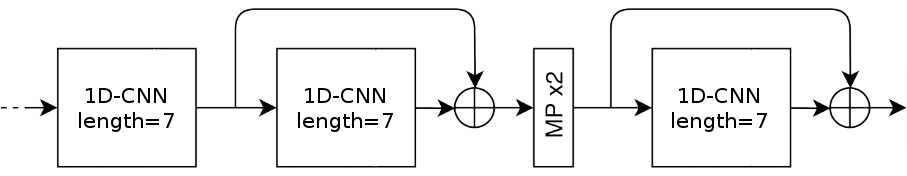}
	\vspace{-7mm}
	\caption{Additional layers for the \textit{frame-level} \& \textit{frame-level many-shapes} architectures, similar as in \cite{dieleman2014end,pons2017end}~-- where MP stands for max pooling.}
	\label{fig:back}
\end{figure}

$\cdot$ \textit{\underline{Frame-level many-shapes:}} consists of a 1D-CNN layer with five filter lengths: 512, 256, 128, 64, 32 \cite{zhu2016learning}. Every filter's stride is of 32 and we use \textit{same} padding -- to easily concatenate feature maps of the same size. Note that out of this single 1D-CNN layer, five feature maps (resulting of the different filter length convolutions) are concatenated. For that reason, every feature map needs to have the same (temporal) size.
Since this model is single-layered and it might be in clear disadvantage when compared to the \textit{sample-level} CNN,
we increase its depth via adding three more 1D-CNN layers (length: 7, stride: 1) -- where the last two layers have residual connections, and the penultimate layer's feature map is down-sampled by two (MP x2), see Figure~\ref{fig:back}.
Averages to construct the feature vector are calculated for each feature map after every 1D-CNN layer.

$\cdot$ \textit{\underline{Frame-level:}} consists of a 1D-CNN layer with a filter of length 512 \cite{dieleman2014end}. Stride is set to be 32 to allow a fair comparison with the \textit{frame-level many-shapes} architecture. As in \textit{frame-level many-shapes}, we increase the depth of the model via adding three more 1D-CNN layers -- as in Figure~\ref{fig:back}. Averages to construct the feature vector are calculated for each feature map after every 1D-CNN layer.

\vspace{-1mm}
\vspace{-2mm}

\noindent\textbf{\textit{Spectrogram inputs}} --- are set to be log-mel spectrograms\linebreak (spectrograms size: 1376$\times$96\footnote{STFT parameters: \textit{window\_size = 512}, \textit{hop\_size=256}, and \textit{fs=12kHz}.}, being $\approx$~29sec of signal).\linebreak Differently from waveform models, spectrogram architectures use no additional layers to deepen single-layered CNNs because these already deliver a reasonable classification performance. Unless we state the contrary, every CNN layer used for processing spectrograms is set to have stride 1. 
As for the \textit{frame-level many-shapes} model, we use \textit{same} padding when many filter shapes are used in the same layer.
The following spectrogram models are studied:

$\cdot$ \textit{\underline{7$\times$96}}: consists of a single 1D-CNN layer with filters of length 7 that convolve through the time axis \cite{dieleman2014end}. As a result: CNN filters are vertical and of shape 7$\times$96. Therefore, these filters can encode spectral (timbral) representations. 
Averages to construct the feature vector are calculated for each feature map after the 1D-CNN layer.

\begin{figure}[h]
	\centering
	\includegraphics[width=1\linewidth]{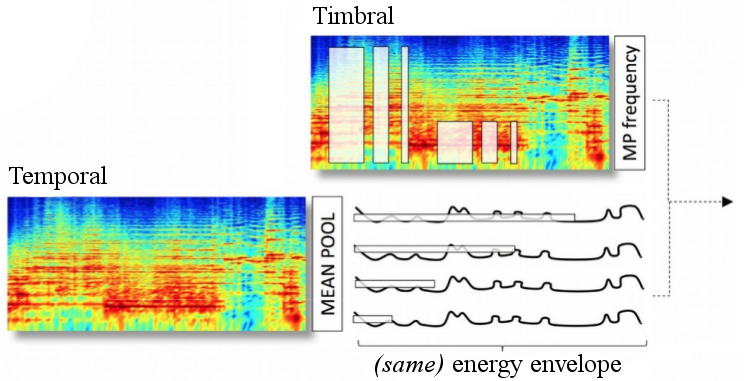}
	\vspace{-8mm}
	\caption{\textit{Timbral+temporal} architecture. MP: max-pool.}
	\label{fig:both}
\end{figure}

$\cdot$ \textit{\underline{7$\times$86}}: consists of a single 2D-CNN layer with vertical filters of shape 7$\times$86 \cite{pons2017timbre,pons2016experimenting}. Since its vertical shape is smaller than the input (86$<$96), filters can also convolve through the frequency axis -- what can be seen as ``pitch shifting" the filter. Consequently, 7$\times$86 filters can encode pitch-invariant timbral representations \cite{pons2017timbre,pons2016experimenting}. 
Further, since the resulting activations can carry pitch-related information, we max-pool the frequency axis to get pitch-invariant features (max-pool shape: 1$\times$11).
Averages to construct the feature vector are calculated for each feature map after the max-pool layer.
\vfill \eject
$\cdot$ \textit{\underline{Timbral}}: consists of a single 2D-CNN layer with many vertical filters of shapes: 7$\times$86, 3$\times$86, 1$\times$86, 7$\times$38, 3$\times$38, 1$\times$38, see Figure \ref{fig:both} (top)\cite{pons2017score,gong2017audio,pons2017timbre}. These filters can also convolve through the frequency axis and therefore, these can encode pitch-invariant representations. Several filter shapes are used to efficiently capture different timbrically relevant time-frequency patterns, e.g.: kick-drums (can be captured with 7$\times$38 filters representing sub-band information for a short period of time),
or string ensemble instruments (can be captured with 1$\times$86 filters representing spectral patterns spread in the frequency axis).
Further, since the resulting activations can carry pitch-related information, we max-pool the frequency axis to get pitch-invariant features (max-pool shapes: 1$\times$11 or 1$\times$59).
Averages to construct the feature vector are calculated for each feature map after the max-pool layer.

$\cdot$ \textit{\underline{Temporal}}: several 1D-CNN filters (of lengths: 165, 128, 64, 32) operate over an energy envelope obtained via mean-pooling the frequency-axis of the spectrogram, see Figure \ref{fig:both} (bottom). 
By computing the energy envelope in that way, we are considering high and low frequencies together while minimizing the computations of the model. Observe that this single-layered 1D-CNN is not operating over a 2D spectrogram, but over a 1D energy envelope~-- therefore no vertical convolutions are performed, only 1D (temporal) convolutions are computed. As seen, domain knowledge 
can also provide guidance to (effectively) minimize the computations of the model. 
Averages to construct the feature vector are calculated for each feature map after the CNN layer.

$\cdot$ \textit{\underline{Timbral+temporal}}: combines both \textit{timbral} and \textit{temporal} CNNs in a single (but wide) layer, see Figure \ref{fig:both}\cite{pons2017end}. \linebreak Averages to construct the feature vector are calculated in the same way as for \textit{timbral} and \textit{temporal} architectures.

$\cdot$ \textit{\underline{VGG}}: is a computer vision model based on a stack of 5 blocks combining 2D-CNN layers (with small rectangular filters of 3$\times$3) and max-pooling (of shapes: 4$\times$2, 4$\times$3, 5$\times$2, 4$\times$2, 4$\times$4, respectively)\cite{choi2016automatic}. 
Averages to construct the feature vector are computed after every pooling layer.
\vspace{-3mm}
\vspace{-7mm}

\noindent As seen, studied architectures are representative of the audio classification state-of-the-art -- introduced in section \ref{sec:arch}. For further details about the models under study: the code is accessible online\footnote{\url{https://github.com/jordipons/elmarc}}, and a graphical conceptualization of the models is available in Figures~\ref{fig:sota}, \ref{fig:back} and \ref{fig:both}.

\vspace{2mm}

\vspace{-2mm}
\subsection{Classifiers: SVM and ELM}
\vspace{-1mm}
We study how several feature vectors (computed considering different CNNs) perform for a given set of classifiers: SVMs and ELMs. We discarded the use of other classifiers since their performance was not competitive when compared to those. SVMs and ELMs are hyper-parameter sensitive, for that reason we perform a grid search:

$\cdot$ \underline{\textit{SVM}} hyper-parameters: we consider both \textit{linear} and \textit{rbf} kernels. For the \textit{rbf} kernel, we set $\gamma$~to: $2^{-3}$, $2^{-5}$, $2^{-7}$, $2^{-9}$, $2^{-11}$, $2^{-13}$, \textit{\#features}$^{-1}$; and for every kernel configuration, we try several \textit{C}'s (penalty parameter): 0.1, 2, 8, 32. We use scikit-learn's SVM implementation~\cite{scikit-learn}.

$\cdot$ \underline{\textit{ELM}}'s main hyper-parameter is the number of hidden units: 100, 250, 500, 1200, 1800, 2500. We use ReLUs as non-linearity, and we use a public ELM implementation.\footnote{\url{https://github.com/zygmuntz/Python-ELM}}

\vspace{-3mm}
\subsection{Datasets: music and acoustic events}
\vspace{-1mm}

\hspace{3mm} $\cdot$  \underline{\textit{GTZAN}} fault-filtered version \cite{tzanetakis2002musical,kereliuk2015deep}. Training songs: 443, validation songs: 197, and test songs: 290 -- divided in 10 classes. We use this dataset to study how randomly weighted CNNs perform for music genre classification. 

$\cdot$  \underline{\textit{Extended Ballroom}} \cite{marchand2016scale, cano2006ismir} -- 4,180 songs divided in 13 classes; 10 stratified folds are randomly generated for cross-validation.
We use this dataset to study how randomly weighted CNNs classify rhythm/tempo classes.

$\cdot$  \underline{\textit{Urban Sound 8K}} \cite{us8k} -- 8732 acoustic events divided in 10 classes; 10 folds are already defined by the dataset authors for cross-validation. Since urban sounds are shorter than 4 seconds and our models accepts $\approx$ 29sec inputs, the signal is repeated to create inputs of the same length.
We use this dataset to study how randomly weighted CNNs perform to classify natural (non-music) sounds.

\vspace{-2mm}
\subsection{Reproducing former results to discuss our method}
\label{sec:repro}

\begin{figure}[h]
	\centering
     \includegraphics[width=1\linewidth]{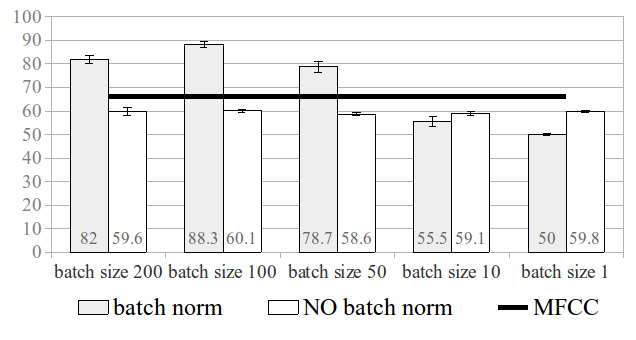}
	\vspace{-9mm}
	\caption{Random CNN features behavior when using (or not) batch normalization. \textit{Dataset}: GTZAN, 10-fold cross-validation. \textit{Performance metric (\%)}: average accuracies (and standard deviations) across 3 runs. \textit{Classifier}: SVM.}
	\label{fig:batch_norm}
\end{figure}

Choi et al. \cite{choi2017transfer} used random CNN features as baseline for their work, and found that (in most cases) these random CNN features perform better than MFCCs. Motivated by this result, we pursue this idea for studying how different deep architectures perform when resolving audio problems.
To start, we first reproduce one of their experiments using random CNNs~-- under the same conditions\footnote{\url{https://github.com/keunwoochoi/transfer_learning_music/} \hspace{1mm}\textit{(more information is available in issue \#2)}}: the GTZAN dataset is split in 10 stratified folds used for cross-validation\footnote{Our work does not use this split, we use the fault-filtered version.}, a VGG architecture with batch normalization is employed, and the classifier is an SVM. We found that results can vary depending on the batch size if, when computing the feature vectors, layers are normalized with the statistics of current batch (batch normalization). For example: if audio-features of the same genre are batch-normalized by the same factor, one can create an artificial \textit{genre cue} that might affect the results. One can observe this phenomena in Figure \ref{fig:batch_norm}, where the best results are achieved when all songs of the same genre fill a full batch (batch size of 100).\footnote{The GTZAN has 10 genres with 100 audios each, one can fill batches of 100 with audios of the same genre via sorting the data by genres.}We also observe that small batch sizes are harming the model's performance~--~see in Figure 5 when batch sizes are set to 1 and 10. And finally, when batch normalization is not used, no matter what batch size we use that the results remain the same -- \textit{ANOVA} test with $H_0$ being that averages are equal\linebreak (\textit{p-value=0.491}). Since it is not desirable that performance depends on the batch size, and that the feature vector of an audio depends on other audios (that are present in the batch): we decided not to use batch normalization.
\vspace{-3mm}

\vspace{-7mm}
\section{Results}
Figures show average accuracies across 3 runs for every feature type (listed on the right with the length of the feature vector in parenthesis). We use a t-test to reveal which models are performing the best -- $H_0$:~averages are equal.
\vspace{-3mm}
\subsection{GTZAN: music genre recognition}
\vspace{-6mm}
\begin{figure}[H]
	\centering
	\includegraphics[width=0.95\linewidth]{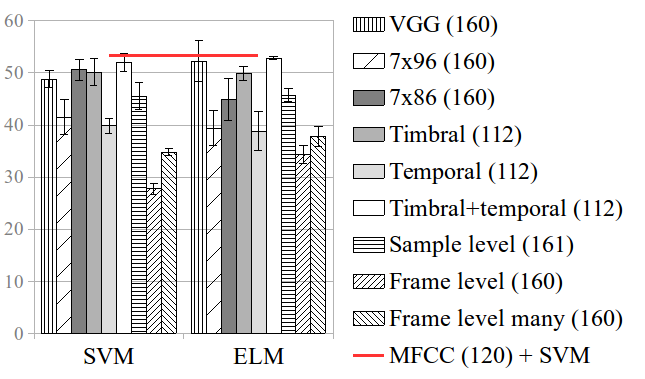}
	\vspace{-4mm}
	\caption{Accuracy (\%) results for the GTZAN dataset with random CNN feature vectors of length $\approx$ 120.}
	\label{fig:parts}
\end{figure}
\vspace{-8mm}
\begin{figure}[H]
	\centering
	\includegraphics[width=0.95\linewidth]{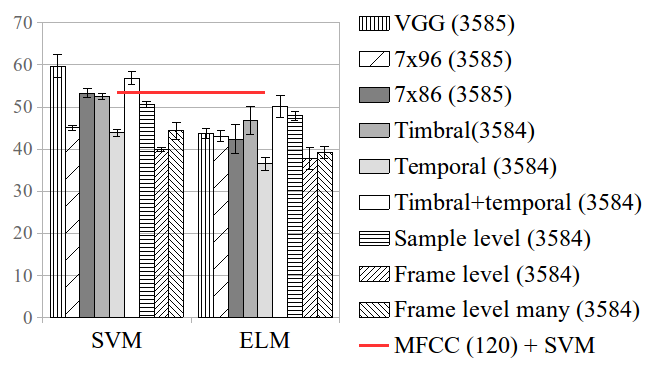}
	\vspace{-4mm}
	\caption{Accuracy (\%) results for the GTZAN dataset with random CNN feature vectors of length $\approx$ 3500.}
	\label{fig:parts}
\end{figure}
\vspace{-3mm}
The \textit{sample-level} waveform model always performs better than \textit{frame-level many-shapes} (t-test: p-value$\ll$0.05).
The two best performing spectrogram-based models are: \textit{timbral+temporal} and \textit{VGG} -- with a remarkable performance of the \textit{timbral} model alone. The \textit{timbral+temporal} CNN performs better than \textit{VGG} when using the ELM~($\approx$3500) classifier (t-test: p-value=0.017); but in other cases, both models perform equivalently (t-test: p-value$>$0.05).
Moreover, the \textit{7x86} model performs better than \textit{7x96} when using SVMs (t-test: p-value$<$0.05), but when using ELMs: \textit{7x96} and \textit{7x86} perform equivalently (t-test: p-value$\gg$0.05).
The best \textit{VGG} and \textit{timbral+temporal} models achieve the following (average) accuracies: 59.65\% and 56.89\%, respectively~-- both with an SVM~($\approx$3500) classifier. Both models outperform the MFCCs baseline: 53.44\% (t-test: p-value$<$0.05), but these random CNNs perform worse than a CNN pre-trained with the Million Song Dataset: 82.1\%~\cite{lee2018samplecnn}.
Finally, note that although \textit{timbral} and \textit{timbral+temporal} models are single-layered, these are able to achieve remarkable performances~-- showing that single-layered spectrogram front-ends (attending to musically relevant contexts) can do a reasonable job without paying the cost of going deep~\cite{pons2017timbre,pons2016experimenting}. Thus meaning, e.g., that the saved capacity can now be employed by the back-end to learn (some more) representations.

\vspace{-3mm}
\subsection{Extended Ballroom: rhythm/tempo classification}
\label{sec:ball}
\vspace{-5mm}
\begin{figure}[H]
	\centering
	\includegraphics[width=0.95\linewidth]{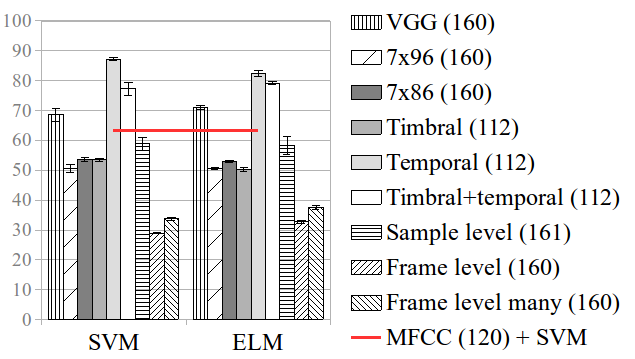}
	\vspace{-4mm}
	\caption{Accuracy (\%) results for the Extended Ballroom dataset with random CNN feature vectors of length $\approx$ 120.}
	\label{fig:parts}
\end{figure}
\vspace{-8mm}
\begin{figure}[H]
	\centering
	\includegraphics[width=0.95\linewidth]{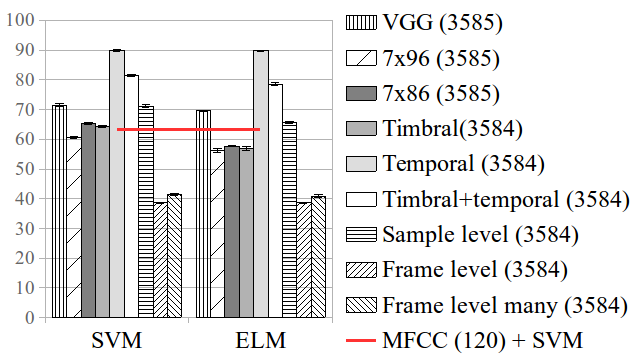}
	\vspace{-4mm}
	\caption{Accuracy (\%) results for the Extended Ballroom dataset with random CNN feature vectors of length $\approx$ 3500}
	\label{fig:parts}
\end{figure}
\vspace{-2mm}
The \textit{sample-level} waveform model always performs better than \textit{frame-level many-shapes} (t-test: p-value$\ll$0.05).
The two best performing spectrogram-based models are: \textit{temporal} and \textit{timbral+temporal}, but the \textit{temporal} model performs better than \textit{timbral+temporal} in all cases (t-test: p-value$\ll$0.05) -- denoting that spectral cues can be a confounding factor for this dataset.
Moreover, the \textit{7x86} model performs better than \textit{7x96} in all cases (t-test: p-value$<$0.05).
The best (average) accuracy score is obtained using \textit{temporal} models and SVMs~($\approx$3500): 89.82\%. Note that the \textit{temporal} model clearly outperforms the MFCCs baseline: 63.25\%\linebreak(t-test: p-value$\ll$0.05) and, interestingly, it performs slightly worse than a trained CNN: 93.7\%~\cite{jeong2017dlr}. This result confirms that the architectures (alone) introduce a strong prior which can significantly affect the performance of an audio model. Thus meaning that, besides learning,
designing effective architectures might be key for resolving (music) audio tasks with deep learning.
In line with that, note that the \textit{temporal} architecture is designed considering musical domain knowledge -- in this case: how tempo \& rhythm are expressed in spectrograms. Hence, its good performance also validates the design strategy of using musically motivated architectures as a way to intuitively navigate through the network parameters space \cite{pons2016experimenting,pons2017timbre,pons2017designing}. 
\vspace{-2mm}
\subsection{Urban Sounds 8K: acoustic event detection}
\vspace{-1mm}
For these experiments we do not use the \textit{temporal} model (with 1D-CNNs of length 165, 128, 64, 32). Instead, we study the \textit{temporal+time} model -- where \textit{time} follows the same design as \textit{temporal} but with filters of length: 64, 32, 16, 8.
This change is motivated by the fact that temporal cues in (natural) sounds are shorter and less important than temporal cues in music (i.e., rhythm or tempo).
\vspace{-4mm}
\begin{figure}[H]
	\centering
	\includegraphics[width=1\linewidth]{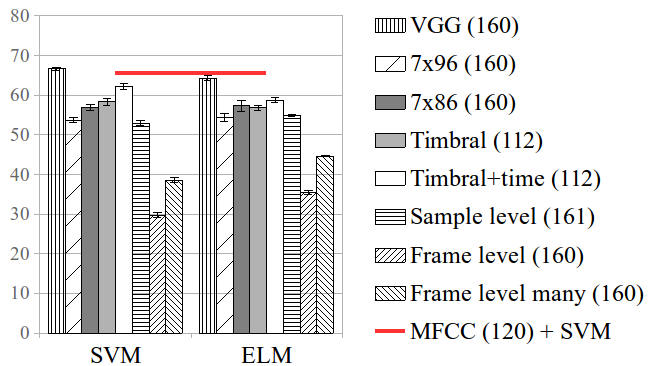}
	\vspace{-6mm}
	\caption{Accuracy (\%) results for the Urban Sounds 8k dataset with random CNN feature vectors of length $\approx$ 120.}
	\label{fig:parts}
\end{figure}
\vspace{-8mm}
\begin{figure}[H]
	\centering
	\includegraphics[width=1\linewidth]{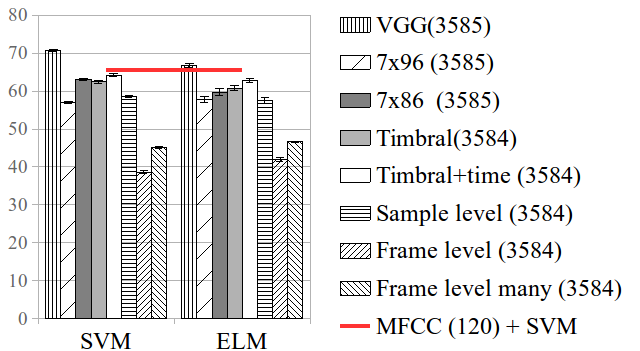}
	\vspace{-7mm}
	\caption{Accuracy (\%) results for the Urban Sounds 8k dataset with random CNN feature vectors of length $\approx$ 3500}
	\label{fig:parts}
\end{figure}
\vspace{-3mm}
\noindent The \textit{sample-level} waveform model always performs better than \textit{frame-level many-shapes} (t-test: p-value$\ll$0.05).
The two best performing spectrogram-based models are: \textit{VGG} and \textit{timbral+time}~-- but \textit{VGG} performs better than \textit{timbral+time} in all cases (t-test: p-value$\ll$0.05).
Also, the \textit{7x86} model performs better than \textit{7x96} in all cases (t-test: p-value$<$0.075).
The best (average) accuracy score is obtained using \textit{VGG} and SVMs ($\approx$3500): 70.74\% -- outperforming the MFCCs baseline: 65.49\%\linebreak(t-test: p-value$<$0.05), and performing slightly worse than a trained CNN: 73\%\footnote{The same CNN achieves 79\% when trained with data augmentation.}\cite{lee2018samplecnn}. Finally, note that \textit{VGG}s achieved remarkable results when recognizing genres and detecting acoustic events -- tasks where timbre is an important cue.
As a result: one could argue that \textit{VGG}s are good at representing spectral features. Hence, these might be of utility for tasks where spectral cues are relevant.

\vspace{-3mm}
\section{CONCLUSIONS}
\vspace{-1mm}
This study builds on top of prior works showing that the (classification) performance delivered by random CNN features is correlated with the results of their end-to-end trained counterparts~\cite{saxe2011random,rosenfeld2018intriguing}. 
We use this property to run a comprehensive evaluation of current deep architectures for (music) audio. Our method is as follows: first, we extract a feature vector from the embeddings of a randomly weighted CNN; and then, we input these features to a classifier -- which can be an SVM or an ELM. 
Our goal is to compare the obtained classification accuracies when using different CNN architectures.
The results we obtain are far from \textit{random}, since: \textit{(i)} randomly weighted CNNs are (in some cases) close to match the accuracies obtained by trained CNNs; and \textit{(ii)}~these are able to outperform MFCCs. This result denotes that the architectures alone are an important piece of the deep learning solution and therefore, searching for efficient architectures capable to encode the specificities of (music) audio signals might help advancing the state of our field.
In line with that, we have shown that (musical) priors embedded in the structure of the model can facilitate capturing useful (temporal) cues for classifying rhythm/tempo classes.
Besides, we show that for waveform front-ends: 
\textit{sample-level} $\gg$ \textit{frame-level many-shapes}~$>$ \textit{frame-level} -- as noted in the (trained) literature \cite{lee2017sample,zhu2016learning,van2016wavenet}.
The differential aspect of the \textit{sample-level} front-end is that its representational power is constructed via hierarchically combining small-context representations, not by exploiting prior knowledge about waveforms.
Further, we show that for spectrogram front-ends:\linebreak\textit{7x96}$<$\textit{7x86} --
as shown in prior (trained) works~\cite{pons2016experimenting,oramas2017multi}. By allowing the filters to convolve through the frequency axis, the architecture itself facilitates capturing pitch-invariant timbral representations.
Finally: \textit{timbral} (\textit{+temporal/time}) and \textit{VGG} spectrogram front-ends achieve remarkable results for tasks where timbre is important -- as previously noted in the (trained) literature \cite{pons2017timbre}.
Their respective advantages being that: \textit{(i)} \textit{timbral} (\textit{+temporal/time}) architectures are single-layered front-ends which explicitly capture acoustically relevant receptive fields -- which can be known via exploiting prior knowledge about the task; and \textit{(ii)} \textit{VGG} front-ends require no prior domain knowledge about the task for its design.
Although our main conclusions are backed by additional results in the (trained) literature, we leave for future work consolidating those via doing a similar study considering trained models.

\section{ACKNOWLEDGMENTS}

This work was partially supported by the Maria de Maeztu Units of Excellence Programme (MDM-2015-0502) -- and we are grateful for the GPUs donated by NVidia.

\bibliography{ISMIRtemplate}

\end{document}